\documentclass[twocolumn,showpacs,preprintnumbers,amsmath,amssymb,superscriptaddress,prx,longbibliography]{revtex4-1}

\usepackage{graphicx}
\usepackage{dcolumn}
\usepackage{bm}
\usepackage{graphics}
\usepackage{subfigure}
\usepackage{graphicx}
\usepackage{epsfig}
\usepackage{epstopdf}
\usepackage{xcolor}
\usepackage[normalem]{ulem}

\allowdisplaybreaks

\begin{document}

\title{Universal relations for dipolar quantum gases}

\author{Johannes Hofmann}
\email{johannes.hofmann@physics.gu.se}
\affiliation{Department of Applied Mathematics and Theoretical Physics, University of Cambridge, 
Centre for Mathematical Sciences, Cambridge CB3 0WA, United Kingdom}
\affiliation{TCM Group, Cavendish Laboratory, University of Cambridge, Cambridge CB3 0HE, United Kingdom}
\affiliation{Department of Physics, Gothenburg University, 41296 Gothenburg, Sweden}

\author{Wilhelm Zwerger}
\email{zwerger@tum.de}
\affiliation{Technische Universit\"at M\"unchen, Physik Department, James-Franck-Strasse, 85748 Garching, Germany}

\date{\today}

\begin{abstract}
We establish that two-dimensional dipolar quantum gases admit a universal description, i.e., their thermodynamic properties 
are independent of details of the interaction at short distances. The only relevant parameters are the dipole length as well as 
the scattering length of the combined short-range plus dipolar interaction potential. We derive adiabatic relations that link the 
change in the thermodynamic potentials with respect to the scattering length and the dipole length to a generalized Tan contact 
parameter and a new dipolar contact, which involves an integral of a short-distance regularized pair distribution function. These 
two quantities determine the scale anomaly in the difference between pressure and energy density and also the internal energy 
in the presence of a harmonic confinement. For a weak transverse confinement, configurations with attractive interactions appear,
which lead to a density-wave instability beyond a critical strength of the dipolar interaction. We show that this instability essentially
coincides with the onset of a roton minimum in the excitation spectrum and may be understood in terms of a quantum analog of the 
Hansen-Verlet criterion for freezing of a classical fluid. \end{abstract}

\maketitle

\section{Introduction}

Interactions in ultracold gases are usually described in terms of the two-body scattering length $a$ as a single parameter, which 
characterizes the complicated and in detail unknown microscopic interaction. Such a reduced description is possible because at 
low energies and densities only two-particle $s$-wave collisions are relevant. Formally, the true interaction is replaced by a zero-range pseudopotential whose strength is adjusted to reproduce a given scattering length~\cite{huang57}. The consequences of 
this description for the associated many-body problem have been elucidated in the independent works by 
Tan~\cite{tan08a,tan08b,tan08c} and by Zhang and Leggett~\cite{zhang09} in the case of two-component Fermi gases with a scattering 
length that is much larger than the typical interaction range $r_e$. They rely on the existence of a well-defined scaling limit, where the 
effective interaction range is taken to zero at a fixed value of the scattering length $a$. In this limit, the detailed form of the combined  
short-range and van der Waals interaction at large separations becomes irrelevant. In particular, the pseudopotential description leads
to set of exact relations for thermodynamic properties like the internal energy or the pressure and also the behavior of the momentum 
distribution $n(q)$ at large wave vectors $q$.  Importantly, since these relations are based on operator identities, they hold for arbitrary 
phases of the many-body problem~\cite{braaten08}. The temperature and the strength of the interaction enter only through a single 
parameter known as the contact ${\cal C}$~\cite{tan08a,tan08b,tan08c}. It is a measure for two atoms to be in close proximity and thus
also determines the amplitude of characteristic power laws that appear in various correlation functions at short distances or times~\cite{braaten12}.

The aim of our present work is to develop an extended universal description for neutral atoms or molecules with
a permanent magnetic or electric dipole moment. Following the realization of a chromium BEC~\cite{griesmaier05}, 
the study of ultracold gases with dipolar interactions has become a major research field, in particular after both Bose and Fermi gases 
of dysprosium~\cite{lu10,lu12} and erbium~\cite{aikawa12,aikawa14} have been cooled into the deeply degenerate regime. In addition, 
stable quantum gases of molecules with a strong electric dipole moment have been created in RbCs~\cite{takekoshi14,molony14} and 
NaK~\cite{park15}. More recently, considerable interest in dipolar gases has been triggered by the observation of supersolid phases, 
in which a periodic density modulation along the weakly-confined direction of a cigar-shaped trap appears in a superfluid state
beyond a critical strength of the dipolar interaction~\cite{boettcher19,tanzi19,chomaz19}. A microscopic description of the 
superfluid-to-supersolid quantum phase transition and its connection to a possible roton instability is still a matter of debate.
It will be shown here that some generic features of this transition can be understood by a generalization of the classical 
Hansen-Verlet criterion for freezing. Moreover, the behavior of the static structure factor at large momentum is determined by
a Tan relation and allows to distinguish the transition to a supersolid from that between a superfluid and a normal, commensurate 
solid which appears in $^4$He. 

Due to the long-range nature of the interaction, an extension of the complete set of Tan relations to 
dipolar gases turns out to be possible only in two dimensions, where the interaction 
\begin{align}
V_d(r) = \frac{d^2}{r^3} \label{eq:Vd}
\end{align} 
for dipoles aligned perpendicular to their motion is purely repulsive in addition to some unknown short-range potential. In this case, 
two-body scattering at low energies is dominated by the s-wave contribution. Moreover, at the many-body level, 
the interaction decays sufficiently fast to give rise to proper thermodynamics with a finite value of the free energy per particle in 
the thermodynamic limit, independent of the boundary conditions~\cite{fisher64}. Neither of these crucial properties holds in three 
dimensions: Here, even for aligned dipoles, the long-range potential $-2d^2\, P_2(\cos{\theta})/r^3$ depends on the angle 
$\theta$ between the direction of alignment and the relative separation ${\bf r}$. As a result, the angular momentum $l$ is not conserved.
The amplitude for angular momentum changing collisions vanishes like $\sim (k\ell_d)^2$ in the 
ultracold limit $k\ell_d\ll 1$~\cite{wang12}, with $\ell_d=md^2/\hbar^2$ the dipolar length. 
However, due to the long-range nature of the interaction, the amplitude
$f(k) \sim \ell_d \ln[ 1/(k\ell_d)]$ for $s$-wave scattering 
diverges at low energies, while the phase shifts $\delta_l(k)=-\tilde{a}_l k$ for finite angular 
momenta start at linear order for arbitrary $l$, with effective 
scattering lengths $\tilde{a}_l\simeq \ell_d/l^2$ that 
decay only slowly with increasing 
$l$~\cite{[][{, Eq. (124.2)}]landau65,bohn09}. Hence, unlike the case of isotropic short-range interactions with a van der Waals tail,
the $s$-wave scattering length is not sufficient to describe the two-body interaction of dipolar gases at low energies in three dimensions 
and no universal description of the thermodynamics and short-distance correlation functions exists.\\

This paper is structured as follows: We begin in Sec.~\ref{sec:2} by considering the strictly two-dimensional dipolar gas where the 
confinement length $l_z$ is much smaller than the dipolar length $\ell_d$. Based on the solution of the associated two-body scattering problem in 
Sec.~\ref{sec:2a}, we derive the short-distance properties of the many-particle wave function in Sec.~\ref{sec:2b} and discuss the 
short-distance properties of the pair distribution, which involves a generalized version of the Tan contact ${\cal C}$. In Section~\ref{sec:2c}, 
we discuss the related adiabatic relations and, in particular,  define a new dipolar analog ${\cal D}$ of the contact parameter. Moreover, 
using an extension to dipolar interactions of an approach due to Fisher and Hohenberg~\cite{fisher88}, explicit results for both contact 
parameters are derived in the low-density limit at zero temperature. In Sec.~\ref{sec:2d}, the two independent contact parameters are 
linked to universal relations for the pressure and the virial theorem. In Sec.~\ref{sec:2e}, we discuss the behavior of the momentum 
distribution and the static structure factor at large wave vectors. Based on previous numerical results, quantitative results for the dipolar 
contact covering the full range of dimensionless coupling constants $\sqrt{n}\ell_d$ are presented in Section~\ref{sec:2f}. In Section~\ref{sec:3}, 
we proceed to discuss the quasi two-dimensional limit, where the motion is restricted to the lowest transverse eigenstate while the associated 
confinement length $l_z$ is still considerably larger than $\ell_d$. In this limit, the form of the interaction potential~\eqref{eq:Vd} holds only at large 
distances, crossing over to an attractive interaction at separations below $l_z$.  The presence of attractively interacting dipoles in this case opens 
the possibility for an instability of the homogeneous superfluid into phases with spatial order. In Section~\ref{sec:3b}, it is shown that the onset of a 
roton minimum in the excitation spectrum coincides with the quantum version of the Hansen-Verlet criterion, which provides an empirical criterion 
for the point at which a fluid phase becomes unstable towards a phase with broken translation symmetry. Moreover, in Section~\ref{sec:3c}, we derive 
exact results for the behavior of the static structure factor at large wave vectors. They allow to distinguish the transition to a supersolid phase that is caused
by partially attractive interactions from the transition of a homogeneous superfluid to a commensurate, non-superfluid crystal, which appears both in 
strictly two-dimensional dipolar gases and also in $^4$He at high pressure due to purely or dominantly repulsive interactions. The paper is concluded 
by a summary in Sec.~\ref{sec:4}. There are three appendices that discuss an example potential that gives rise to universal dipolar scattering, 
derive the adiabatic relations,  and generalize the universal relations in two dimensions to general repulsive power-law interactions. 

\section{Universal relations for strictly two-dimensional dipolar gases}\label{sec:2}
 
In this section, we derive universal relations for a Bose gas with dipolar interactions in the limit where scattering is
purely two-dimensional in nature. The many-body Hamiltonian of such a system with $N$ particles is 
\begin{align}
\hat{H} &= - \frac{\hbar^2}{2m} \sum_{i=1}^N \nabla_i^2 + \sum_{i<j} V({\bf r}_i - {\bf r}_j) .
\label{eq:hamiltonian}
\end{align}
Here, $V({\bf r})$ is the complete effective interaction as truncated to motion in the plane.  For separations $r>R_e$ larger than a short-distance range, 
the potential is assumed to be of the pure dipolar form given in Eq.~\eqref{eq:Vd} while for $r < R_e$, it changes to an unknown short-range form. 
As discussed by~\textcite{buechler07}, the short-distance cutoff in the presence of a transverse confinement with oscillator length $l_z$ is of order 
$R_e\simeq (l_z/\ell_d)^{4/5} \ell_d$. The existence of a proper zero-range scaling limit thus requires a tight transverse confinement 
with an associated length $l_z$ considerably smaller than the dipolar length $\ell_d$. In this limit, both the scattering length $a_2$ associated with 
the full two-body interaction $V({\bf r})$ and also the dipolar length $\ell_d$ are much larger than $R_e$ (note that the scattering length in two 
dimensions is denoted by a subscript $a_2$). 

Throughout the paper, we consider a system of bosons, where only s-wave scattering is relevant at low energies. However, our results 
on the short-distance correlations and their connections to thermodynamic properties also hold for two-component Fermi gases with 
only minor modifications by factors of two. In fact, the latter problem is of relevance not 
only in the context of ultracold atoms but also arises in two-dimensional electron gases (2DEG).  As discussed by~\textcite{spivak04}, 
a realization of a two-component Fermi gas with dipolar interactions is provided by a 2DEG in a MOSFET device with a ground plane 
at a distance $\tilde{d}$. The interaction between the electrons in the 2DEG is then of a pure dipolar form at distances larger than $\tilde{d}$ 
with an effective dipole moment $d^2_{\rm eff}=4e^2 \tilde{d}^2/\epsilon$, where $\epsilon$ is the dielectric constant of the host semiconductor.  
Since the electrons are in an equal mixture of spin-up and spin-down state, scattering appears both in a relative singlet state associated 
with even angular momenta $m$ or in relative triplet states, which involves odd $m$.
At low densities, where the Fermi wave vector $k_F$ obeys $\ln(1/k_F\ell_d^{\rm eff})\gg k_F\ell_d^{\rm eff}$, the $s$-wave 
contribution dominates and thus electron-electron interactions in a relative triplet state become irrelevant. More generally, however, 
the scattering phase shifts $\delta_m(k) \sim k \ell_d$ associated with the long-range dipolar interaction in two dimensions are of the 
same order for arbitrary finite angular momenta $m\neq 0$~\cite{ticknor09,friedrich13}. A proper discussion of the electronic many-body 
problem at realistic densities, where $k_F\ell_d^{\rm eff}\simeq (4/r_s)\, (\tilde{d}/a_B)^2$ is much larger than one, thus requires to 
include all possible values of $m$. Here, $a_B = \hbar^2 \varepsilon/me^2$ is the effective Bohr radius and $r_s a_B= 1/\sqrt{\pi n}$~\cite{vignale05}. 

\subsection{Two-body scattering}\label{sec:2a}

Before turning to the full many-body problem, we consider the two-body problem with the pure dipolar potential $V_d(r)$. 
The scattering wave function may be expanded in partial waves as $\psi({\bf r}) = \frac{1}{\sqrt{\tilde{r}}} \sum_m e^{im\varphi} \phi_m(\tilde{r})$, 
where $m\in\mathbb{Z}$ is the integer angular momentum and $\varphi$ the angle in the 2D plane. Introducing $\tilde{r} = r/\ell_d$ as the 
dimensionless radius and $\tilde{k} = k \ell_d$ as the corresponding relative wave vector, the Schr\"odinger equation for the relative motion reads
\begin{align}
\biggl[- \frac{d^2}{d \tilde{r}^2} + \frac{m^2 - \frac{1}{4}}{\tilde{r}^2} + \frac{1}{\tilde{r}^3} - \tilde{k}^2\biggr] \phi_m(\tilde{r}) &= 0 . \label{eq:dimensionless}
\end{align}
The low-energy scattering properties are dominated by the $s$-wave solution. The associated 
scattering-phase shift $\delta_0(k)$ has a logarithmic dependence on momentum, which is characteristic for short-range interactions in two 
dimensions~\cite{adhikari86}. It is parametrized by a scattering length $a_2$ defined by the dominant first term in the expansion
\begin{equation}
\cot \delta_0(k) = \frac{2}{\pi} \ln \frac{ka_2 e^{\gamma_E}}{2} - \frac{4 \alpha}{\pi} (k\ell_d) \ln^2 ka_2 + {\cal O}(k) ,
\label{eq:delta}
\end{equation}
where $\gamma_E \approx 0.577$ is the Euler constant and $\alpha$ is a positive numerical factor of order one. Note that for a dipolar interaction in two dimensions, 
the standard effective-range expansion does not hold~\cite{arnecke08}. The subleading 
term in the scattering phase shift~\eqref{eq:delta} is therefore not of the usual form $\sim\!k^2$ but is nonanalytic~$\sim\!|k|$ in the momentum 
and also contains an additional logarithmic factor~\footnote{The breakdown of the effective-range expansion is apparent from a divergence in the Bethe integral expression for the scattering 
phase shift when evaluated using the dipolar two-body wave function at threshold~\eqref{eq:2bodywf}~\cite{bethe49,arnecke08}. The correction 
in Eq.~\eqref{eq:delta} is obtained by regulating this divergence with an upper cutoff of order $1/k$.}. 
The numerical factor $\exp{(\gamma_E)}/2$ in the leading contribution has been chosen such that no 
corrections of order $k^0$ are present. While this factor is often absorbed in the definition of $a_2$ (see, for example, Ref.~\cite{bloch08}), 
our convention for the phase shift~\eqref{eq:delta} ensures that the asymptotic form $\lim_{r\to\infty} \lim_{k\to0} \phi(r) \sim \ln (r/a_2)$ of 
the two-body wave function at zero energy has no corrections of order ${\cal O}(r^0)$.
The scattering states $\phi_m$ at small energies can be determined analytically in terms of modified Bessel functions~\cite{ticknor09}. 
For a pure dipolar interaction, one thus obtains $a_{2}^d = e^{2\gamma_E} \ell_d$~\cite{ticknor09}, 
as mentioned above. In order to deal with the realistic situation of an additional short-range part of the interaction at distances below 
the potential range $R_e$, it is sufficient to add the irregular 
solution in a pure dipolar potential with a prefactor $\ln (a_2/a_{2}^d)$. As a result, the two-body wave function is of the form 
\begin{align}
\phi(r) &= 2 K_{0}\Bigl(\sqrt{\tfrac{4 \ell_d}{r}}\Bigr) - \ln\Bigl( \tfrac{a_2}{a_{2}^d}\Bigr) \, I_{0}\Bigl(\sqrt{\tfrac{4 \ell_d}{r}}\Bigr) . \label{eq:2bodywf}
\end{align}
An example of a potential that gives rise to this universal scattering form is discussed in App.~\ref{app:B}. At short distances $r\ll \ell_d$, 
the regular and irregular parts scale as an exponential with $(r/\ell_d)^{1/4} \exp[{\mp \sqrt{4 \ell_d/r}}]$, 
which follows from a WKB approximation~\cite{frank71}. An important point to note is that the singular behavior $\sim 1/r^n$ with $n > 2$ 
of the dipolar potential at short distances dominates both the kinetic energy and the angular momentum contribution~\cite{case50,frank71}.
As a result, the physically meaningful and relevant parameter is the full scattering length $a_2$ of the combined short-range plus dipolar 
potential in addition to the dipolar length $\ell_d$. 

\subsection{Pair distribution function and contact}\label{sec:2b}

A systematic method that connects the short-distance properties of a many-body system and two-body wave functions in vacuum is the 
operator product expansion (OPE)~\cite{braaten08,hofmann11,braaten12,goldberger12}. For power-law interactions, this technique has 
been used previously in the Coulomb problem in Ref.~\cite{hofmann13}. Here, we follow a more intuitive approach that relies on the 
short-distance factorization of many-body wave functions. Specifically, we consider the pair distribution function $g(r)$, which describes 
the probability density of detecting pairs of particles separated by a distance ${\bf r}$. Its formal expression
\begin{align}
g({\bf R}, {\bf r}) &= \frac{N(N-1)}{n^2} \int d{\bf X} \, |\Psi({\bf R} -\frac{{\bf r}}{2}, {\bf R} + \frac{{\bf r}}{2}, {\bf X})|^2 \label{eq:Cdd}
\end{align}
contains the $N$-particle wave function $\Psi$ in position space, where we introduce the short-hand ${\bf X} = ({\bf r}_3, \ldots, {\bf r}_N)$ 
for coordinates that are integrated over.  We also use relative and center-of-mass coordinates ${\bf r}$ and ${\bf R}$ for the first two 
particle coordinates. For a homogeneous system, there will be no dependence on ${\bf R}$.  The basic assumption, which can be proven formally 
within the OPE, is that whenever two particle coordinates are close to each other, the wave function should approach the relative two-body solution 
up to a factor that remains finite at ${\bf r}=0$. The many-body wave function thus factorizes according to 
\begin{align}
\lim_{{\bf r}_1 \to {\bf r}_2} \Psi({\bf r}_1, {\bf r}_2, {\bf X}) &= \phi({\bf r}) {\cal A}({\bf R}; {\bf X}) , \label{eq:manybody}
\end{align}
where $\phi({\bf r})$ is defined in Eq.~\eqref{eq:2bodywf} and ${\cal A}({\bf R}; {\bf X})$ is a remainder that does not depend on the relative 
coordinate~${\bf r}$. Using Eq.~\eqref{eq:manybody} in the definition of the pair distribution function gives the universal short-distance behavior
\begin{align}
\lim_{r\to 0} n^2 g({\bf R}, {\bf r}) &= 
\frac{|\phi(r)|^2}{(2 \pi)^2} {\cal C}({\bf R}) , \label{eq:shortrangegr}
\end{align}
where we introduce the contact density ${\cal C}({\bf R})$
\begin{align}
{\cal C}({\bf R}) &= (2 \pi)^2 \, N(N-1) \int d{\bf X} \, |{\cal A}({\bf R}; {\bf X})|^2 . \label{eq:contactdensity}
\end{align}

\begin{figure}[t]
\scalebox{0.9}{\includegraphics{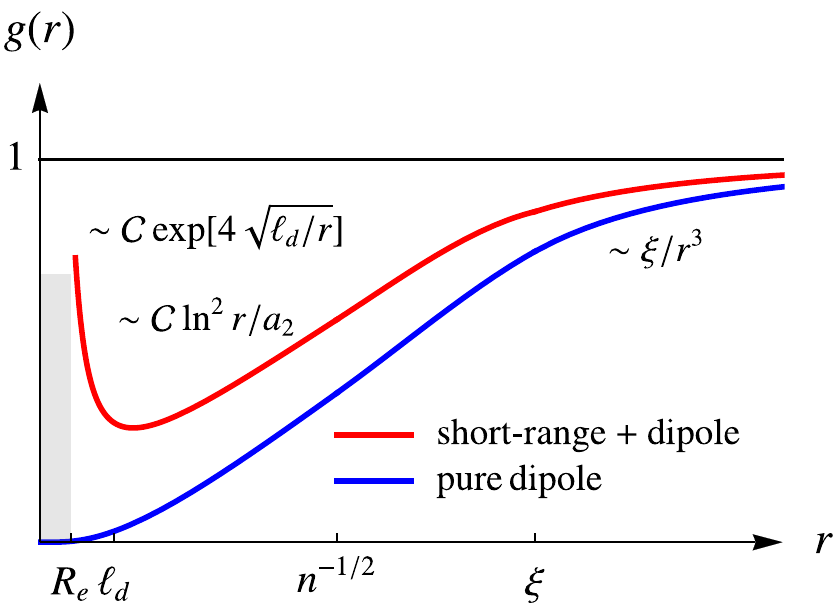}}
\caption{Schematic plot of the pair distribution function for a system with combined dipole and short-range interaction (red line). 
For comparison, we also include the restricted case of a pure dipole interaction without a short-range part, for which $a_2 = e^{2\gamma_E} \ell_d$ (blue line). 
For distances that are small compared to the inter-particle separation $1/\sqrt{n}$ but still larger than the potential range $R_e$, the pair distribution 
function is universal [Eq.~\eqref{eq:shortrangegr}], with a magnitude set by the contact~\eqref{eq:contactdensity}.}
\label{fig:1}
\end{figure}

For a pure dipolar interaction $a_2=a_{2}^d$, the pair distribution function is exponentially suppressed for $r \ll \ell_d$. In the actual relevant 
situation of an interaction which differs from the $1/r^3$-behavior at short distances, however, $g(r)$ diverges exponentially near the origin, 
which is a consequence of the presence of the irregular solution $I_0$ in Eq.~\eqref{eq:2bodywf}. The resulting overall form of the pair 
distribution  function is illustrated in Fig.~\ref{fig:1}, where the continuous blue line denotes the case of a pure dipole 
interaction, while the red line qualitatively describes the realistic situation of a combined dipole and short-range interaction.
In observables that involve an integral over the pair distribution function, the exponential divergence of $g(r)$ at short distances must of 
course be canceled, as will be shown explicitly in Eq.~\eqref{eq:D} below. In the limit of separations $r\gg \ell_d $ but still much smaller 
than the average interparticle distance, the pair distribution function exhibits a logarithmic dependence $g(r) \sim \ln^2 (r/a_2)$, 
which is the standard result for a two-dimensional system with finite scattering length~\cite{werner12a,werner12b}. Note that this 
latter regime only exists in the low-density limit $\sqrt{n}\ell_d \ll 1$, where the dipole length is much smaller than the interparticle separation. 

As indicated in Fig.~\ref{fig:1}, the pair distribution function has a universal form also at large distances: At zero temperature and for any 
compressible fluid phase, it approaches the asymptotic value one from below with an inverse cube power law
\begin{align}
\lim_{r \to \infty} g(r) &= 1 - \frac{\xi}{2\pi\sqrt{2}\, n r^3} + \ldots  \label{eq:grlarger}
\end{align}  
This dependence is a consequence of the nonanalytic behavior $S(q\to 0)=\vert\mathbf{q}\vert\,\xi/\sqrt{2}+\ldots$ of the static structure factor 
at small momentum that defines the characteristic length $\xi$. The large-distance result~\eqref{eq:grlarger} then follows using the standard connection 
\begin{align}
S({\bf q}) &= 1 + n \int d{\bf r} \, e^{-i{\bf q} \cdot {\bf r}} (g(r) - 1)  \label{eq:defSq}
\end{align}
between the static structure factor and the pair distribution function. A quite general upper bound on the length $\xi$ has been derived by Price~\cite{price54} 
using a combination of the $f$-sum rule and the compressibility sum rule. Defining $\tilde{\kappa}=\partial n/\partial\mu$ as an effective compressibility, it reads
\begin{align}
\xi \leq \hbar \sqrt{\frac{\tilde{\kappa}}{2mn}} . \label{eq:boundprice}
\end{align}
Within a single-mode approximation, where the density fluctuation spectrum 
at long wavelengths is exhausted by a single collective mode, the bound becomes an equality. In this case the length $\xi = \hbar/\sqrt{2}mc_s$ 
is uniquely determined by the speed of sound $c_s$.

\subsection{Adiabatic relation}\label{sec:2c}

In the zero-range limit $a_2,\ell_d\gg R_e$, all information on the interaction is contained in the total scattering length $a_2$ and 
the dipolar length $\ell_d$. One may therefore consider the change ($A$  denotes the area of the system)
\begin{align}
d\Omega &= - S dT - P dA - N d\mu + X_a \, d(\ln a_2) + X_d \, d(\ln \ell_d) , \label{eq:defcontacts}
\end{align}
of the grand canonical potential $\Omega=-P A$ in response to changes of these two parameters, which defines two extensive quantities 
$X_a$ and $X_d$. They are the generalized forces conjugate to the variables $\ln a_2$ and $\ln \ell_d$ in $\Omega(T,A,\mu,a_2,\ell_d)$.
In physical terms, $X_a$ and $X_d$ describe the work done on the system under changes in the scattering length or the dipole length at 
fixed temperature $T$, area $A$, and chemical potential $\mu$. As shown in App.~\ref{app:derivation}, these forces are related to the contact 
parameter defined in Eq.~\eqref{eq:contactdensity} in the following manner:
\begin{align}
\frac{X_a}{A} &= \frac{\partial \varepsilon}{\partial (\ln a_2)} \biggr|_{\ell_d} = \frac{\hbar^2}{4 \pi m} {\cal C} \label{eq:adiabatic1} \\
\frac{X_d}{A} &= \frac{\partial \varepsilon}{\partial (\ln \ell_d)} \biggr|_{a_2} = {\cal D}_{} \, ,
\label{eq:adiabatic2}
\end{align}
where the partial derivatives of the energy density $\varepsilon=E/A$ are taken at fixed dipole length and scattering length, respectively, as well as
at fixed entropy $S$, particle number $N$, and area $A$, and where the quantity ${\cal D}$ is defined in terms of the pair distribution function as follows
\begin{align}
{\cal D}_{}  
&= \frac{d^2}{2} \int d{\bf r} \, \frac{n^2 g(r) - \frac{|\phi(r)|^2}{(2 \pi)^2} {\cal C}}{r^3} . \label{eq:D}
\end{align}
The first expression~\eqref{eq:adiabatic1} is the standard Tan adiabatic theorem in two dimensions, generalized to the situation 
where the scattering length also includes the contribution from the dipolar interaction. It establishes the fact that the contact is finite 
and positive also in the presence of dipolar interactions. As a result, the energy is an increasing function of the scattering length~\cite{tan08a}. 
The second relation is new and specific for gases with dipolar interactions. More generally, as shown in App.~\ref{app:general}, Eq.~\eqref{eq:D}  
may be extended to repulsive inverse power law potentials of the form $1/r^{2+\sigma}$ with arbitrary values $\sigma >0$.
Eq.~\eqref{eq:adiabatic2} defines a dipolar analog ${\cal D}$ of the contact, which may be negative in general as shown below but is always finite.  
Indeed, the second term in the integral in Eq.~\eqref{eq:D} precisely cancels the short-distance divergence of the 
pair distribution function, Eq.~\eqref{eq:shortrangegr}, and thus renders the expression finite and independent of short-distance details. 
As will be shown in Eq.~\eqref{eq:pressure} below, the dipolar contact ${\cal D}$ may be understood as a nonanomalous contribution measuring the 
deviation in the difference $P-\varepsilon$ between pressure and energy density due to the fact that a $1/r^3$-interaction violates 
scale invariance explicitly. The adiabatic relations~\eqref{eq:adiabatic1} and~\eqref{eq:adiabatic2} are stated for a homogenous system, 
however the extension to inhomogeneous or few-body states is straightforward. In this case, the pair distribution function $g({\bf R}, {\bf r})$ 
and both contact densities ${\cal C}({\bf R})$ as well as ${\cal D}({\bf R})$ depend on the center-of-mass coordinate ${\bf R}$. 
Upon integration over ${\bf R}$, they give rise to extensive values of $X_a$ and $X_d$.  

For a vanishing dipolar interaction, the full scattering length $a_2$ reduces to the two-dimensional scattering length of the short-range potential. 
The first adiabatic relation~\eqref{eq:adiabatic1} then coincides with the standard adiabatic relation for bosons with short-range interactions~\cite{werner12b}. 
Since the pair distribution function in this case  behaves as
\begin{equation}
n^2 g(r)\vert_{d=0} = \frac{\ln^2(r/a_2)}{(2 \pi)^2}\, {\cal C} + {\cal O}(r^2) 
\label{eq:grshortd=0}
\end{equation}
at short distances, the integral in Eq.~\eqref{eq:D} converges. The dipolar contact density ${\cal D}$ thus vanishes as ${\cal O}(d^2)$ 
(with logarithmic corrections in $na_2^2$, see below), as expected. In the opposite limit of a negligible short-range contribution,  the 
scattering length $a_2=a_{2}^d \simeq \ell_d$ is fixed at the value obtained for a pure dipolar interaction. It is then no longer an independent 
thermodynamic variable separate from $\ell_d$. As a result, the derivative with respect to $\ell_d$ gives a single adiabatic relation that is the 
sum of~\eqref{eq:adiabatic1} and~\eqref{eq:adiabatic2}. Using the fact that the second term in the integral of Eq.~\eqref{eq:D} is finite for a 
pure dipolar interaction and cancels the contact term stemming from Eq.~\eqref{eq:adiabatic1}, one obtains
\begin{align}
\tilde{\cal D} &= \frac{\partial \varepsilon}{\partial (\ln \ell_d)} \biggr|_{a_2=a_{2}^d}
= \frac{d^2}{2} \int d{\bf r} \, \frac{n^2 g(r) }{r^3} ,\label{eq:dipoleadiabatic}
\end{align}
which is just the interaction energy density. This result can also be obtained directly using the Hellmann-Feynman theorem.

As emphasized above, the thermodynamic relations~\eqref{eq:adiabatic1} and~\eqref{eq:adiabatic2} hold for arbitrary states of the many-body problem, 
both at vanishing and at finite temperature. The calculation of the associated contact coefficients requires, however, a quantitative solution of the many-body 
problem, which is in general possible only numerically. Explicit results can be derived at zero temperature and low densities and also in the 
nondegenerate limit by means of a virial expansion, following the approach in Ref.~\cite{barth15} for the two- and three-body contacts of Bose gases 
with short-range interactions in three dimensions. In order to determine the contact densities ${\cal C}$ and ${\cal D}$ in the ground state 
at low densities, we use an approach due to Fisher and Hohenberg~\cite{fisher88}. Based on results by Popov~\cite{popov72}, they showed 
that for a quite general form of the two-body interaction, the dependence $\mu(n)$ of the chemical potential on the density $n$ at small densities 
can be obtained from a perturbative solution of the implicit equation
\begin{align}
\mu = n \vert t(0,0,E=\mu)\vert=\frac{4\hbar^2n/m}{|\cot\delta_0(k=\sqrt{m\mu/\hbar^2}) - i |} ,
\label{eq:T-matrix}
\end{align}
which involves the two-body T-matrix at vanishing total momentum evaluated at a finite energy $E=\mu$ in the center-of-mass frame. 

In the presence of long-range dipolar interactions, the effective range expansion of the scattering 
phase shift in Eq.~\eqref{eq:delta} gives rise to an equation of state at low densities of the form 
\begin{align}
n(\mu) &= \frac{m\mu}{4\pi\hbar^2} \bigg\{\frac{1}{\varepsilon(\mu)} + \frac{8 I }{\pi} \varepsilon(\mu) \biggl(1 + \frac{\varepsilon(\mu)}{2}\biggr) \nonumber \\*
&\qquad + \alpha \ell_d \sqrt{\frac{m\mu}{\hbar^2}} \frac{1}{\varepsilon^2(\mu)} + \ldots\biggr\} , \label{eq:EOS}
\end{align}
where $\varepsilon^{-1}(\mu) = \ln[4\hbar^2/(m\mu a_2^2 e^{2\gamma_E+1})]$.  The first term is the universal result for the chemical potential 
of a Bose gas with scattering length~$a_2$~\cite{schick71,fisher88}. In addition, we also include the leading and universal logarithmic corrections 
in the small parameter $\varepsilon(\mu)\ll 1$  with $I=1.0005$ a numerical constant, which were determined by~\textcite{mora09}. These corrections
are not contained in the Fisher-Hohenberg ansatz~\eqref{eq:T-matrix}. In the ultra-low density limit, where $\varepsilon(\mu)\ll 1$, the equation of state 
only depends on the low-energy scattering length $a_2$~\cite{astrakharchik09}, and there is no separate dependence on the dipole length~$\ell_d$. 
The correction in Eq.~\eqref{eq:EOS} proportional to $\sqrt{n}\ell_d$  arises from the long-range nature of the dipolar interaction 
and becomes relevant beyond the limit of ultra-low densities. In particular, for a pure dipolar gas with $a_2 = a_2^d$, this term is larger than 
the Castin-Mora correction for $m\mu a_2^2/\hbar^2 \gtrsim 2 \cdot 10^{-4}$. This corresponds to densities $\sqrt{n}\ell_d \gtrsim 10^{-3}$,
which covers the relevant range in Fig.~\ref{fig:3} below. The fact that the equation of state of a 2D Bose gas with dipolar interactions
differs substantially from the case of short-range interactions in this regime of densities has indeed been observed in numerical calculations 
by~\textcite{astrakharchik09}. The contact parameter can be determined from Eq.~\eqref{eq:T-matrix} or~\eqref{eq:EOS} using
\begin{equation}
n d\mu =dP- sdT + \frac{\hbar^2}{4\pi m} {\cal C}\, d(\ln a_2) + {\cal D}\, d(\ln \ell_d) ,
\label{eq:Gibbs-Duhem}
\end{equation}
which follows from the definition of the contacts~\eqref{eq:defcontacts} and the Gibbs-Duhem relation. At low densities, the resulting contact 
\begin{align}
{\cal C}(a_2) &= \bigg(\frac{4 \pi n}{\ln \left[4e^{-2\gamma_E}/(na_2^2)\right] }\biggr)^2 + \ldots \label{eq:contactpert}
\end{align}
only involves the scattering length and ---  apart from the logarithmic factor --- essentially vanishes with the square of the density. 
The dipolar contact arises from the contribution $\sim\ell_d$ in Eq.~\eqref{eq:EOS} and is given by 
\begin{align}
{\cal D}&= - \frac{16\hbar^2 \alpha}{5m} \frac{\pi^{3/2} n^{5/2}\ell_d}{\ln^{1/2} [4e^{-2\gamma_E}/(na_2^2)]} + \ldots . \label{eq:dcontactpert}
\end{align}
It is negative but vanishes faster with density than ${\cal C}$. A similar behavior is found in the two-body limit,
where a bound state of the combined short-range plus dipolar interaction exists whenever $a_2>a_2^d$~\footnote{J. Hofmann and W. Zwerger, unpublished}. 
Using the adiabatic relation $\partial E/\partial (\ln{a_2})=\hbar^2 C/(4\pi m)$, the resulting (integrated) contact $C_{\rm 2-body}=32\pi e^{-2\gamma_E}/a_2^2$ is 
positive while $D_{\rm 2-body}=-{\rm const} \cdot \hbar^2/(m\ell_d^3)$ is negative with a numerical prefactor ${\rm const}$ of order one.    
In the special case of a pure dipole interaction, where $a_2 = a_{2}^d \sim \ell_d$, 
only the dipolar contact remains. At ultra-low densities, it is determined by the perturbative result~\eqref{eq:contactpert} 
apart from a trivial factor, i.e.,
\begin{align}
\tilde{\cal D} &= \frac{\hbar^2}{4 \pi m} {\cal C}(a_2=a_{2}^d) . \label{eq:puredipole}
\end{align}

\subsection{Pressure relation and virial theorem}\label{sec:2d}

The adiabatic relations give rise to two additional exact expressions for the pressure in a uniform
situation or the total energy in a harmonic trap.  
First, we derive the pressure relations for a uniform gas. To this end, we write the grand canonical potential in dimensionless form 
\begin{align}
&\Omega(T, \mu, A, 1/a, 1/\ell_d) \nonumber \\
&= k_B T \, \tilde{\Omega}(\frac{\mu}{k_B T}, \frac{\hbar^2/mA}{k_BT}, \frac{\hbar^2/ma_{2}^2}{k_BT}, \frac{\hbar^2/m\ell_d^2}{k_BT}) .
\end{align}
This relation implies the scaling law
\begin{align}
\Omega(\lambda T, \lambda \mu, A/\lambda, \sqrt{\lambda}/a_2, \sqrt{\lambda}/\ell_d) = \lambda \Omega(T, \mu, A, 1/a_2, 1/\ell_d) .
\end{align}
Taking the derivative of this expression with respect to $\lambda$ and evaluating the result at $\lambda = 1$ gives 
\begin{align}
(T \frac{\partial}{\partial T} + \mu \frac{\partial}{\partial \mu} - A \frac{\partial}{\partial A} - \frac{1}{2} \frac{\partial}{\partial (\ln a_2)} - 
\frac{1}{2} \frac{\partial}{\partial (\ln \ell_d)}) \Omega = \Omega .
\end{align}
Using $\bigl(T \frac{\partial}{\partial T} + \mu \frac{\partial}{\partial \mu} - A \frac{\partial}{\partial A} - \Omega\bigr) \Omega
 = - T S - \mu N + P A - \Omega = P A - E$, we obtain a relation for the pressure:
 \begin{align}
 P &= \varepsilon + \frac{\hbar^2 {\cal C}}{8 \pi m} + \frac{{\cal D} }{2} . \label{eq:pressure}
 \end{align}
For vanishing dipolar strength $d^2 \to 0$, the expression reduces to the 2D Tan relation~\cite{tan08c,valiente11,hofmann12}
\begin{align}
P(\ell_d=0) = \varepsilon + \frac{\hbar^2 {\cal C}}{8 \pi m} . \label{eq:pressuretan}
\end{align} 
As noted above, the remaining contact term $\sim {\cal C}$ arises as an anomaly due to the fact that a zero-range interaction in two 
dimensions is scale invariant only at the classical level. The invariance is broken in the quantum theory where the coupling constant 
becomes scale-dependent~\cite{hofmann12}. In the opposite limit of a dipolar interaction without a short-range part, the result 
\begin{align}
P(a_2=a_{2}^d) &= \varepsilon + \frac{\tilde{\cal D}}{2}  \label{eq:pressurepure}
\end{align}
is equivalent to the virial theorem for a pure power law interaction $\sim 1/r^3$ since, as pointed out in Eq.~\eqref{eq:dipoleadiabatic}, 
$\tilde{\cal D}$ is just the interaction energy density.  

A different version of the virial theorem can be derived for dipolar gases that are confined by a harmonic radial trapping potential 
 $V_{\rm ext}({\bf r}) = m\omega^2r^2/2$ with frequency $\omega$ (for simplicity, we assume an isotropic trap, however the final 
result holds also in the anisotropic case). The associated grand canonical potential can then be written in dimensionless form as
\begin{align}
&\Omega(T, \mu, \omega, 1/a, 1/\ell_d) \nonumber \\*
&= k_B T \, \tilde{\Omega}(\frac{\mu}{k_B T}, \frac{\hbar \omega}{k_BT}, \frac{\hbar^2/ma_2^2}{k_BT}, \frac{\hbar^2/m\ell_d^2}{k_BT}) .
\end{align} 
 A similar scaling analysis as above gives 
 \begin{align}
 (T \frac{\partial}{\partial T} + \mu \frac{\partial}{\partial \mu} + \omega \frac{\partial}{\partial \omega} - 
 \frac{1}{2} \frac{\partial}{\partial (\ln a_2)} - \frac{1}{2} \frac{\partial}{\partial (\ln \ell_d)}) \Omega = \Omega .
\end{align}  
 Using $\bigl(T \frac{\partial}{\partial T} + \mu \frac{\partial}{\partial \mu} - 1\bigr) \Omega = - T S - \mu N - \Omega = - E$ and that the partial 
 derivative of the grand canonical potential with respect to the trapping frequency is equal to the derivative of the energy (at fixed entropy), we obtain 
 \begin{align}
 (\omega \frac{\partial}{\partial \omega} - \frac{1}{2} \frac{\partial}{\partial (\ln a_2)} - \frac{1}{2} \frac{\partial}{\partial (\ln \ell_d)}) E = E .
 \end{align}
 Now, using $\omega \partial E/\partial \omega = 2 \langle V_{\rm ext} \rangle$, we obtain the virial theorem
 \begin{align}
 E &= 2 \langle V_{\rm ext} \rangle - \frac{\hbar^2}{8 \pi m} \int_{\bf R} \, {\cal C}({\bf R}) - \frac{1}{2} \int_{\bf R} \, {\cal D}
 ({\bf R})  . \label{eq:virialfull}
 \end{align}
 Again, the first two terms are the standard virial theorem for 2D quantum gases~\cite{tan08c,werner08,valiente11,hofmann12}. 
 For a pure dipolar interaction, the virial theorem reduces to
  \begin{align}
 E &= 2 \langle V_{\rm ext} \rangle - \frac{1}{2} \int_{\bf R} \, \tilde{\cal D}
 ({\bf R}) . \label{eq:virialpure}
 \end{align}
 This result can be compared with a virial theorem by~\textcite{goral01} that was derived for trapped single-component dipolar 
 Fermi gases in three dimensions in the semi-classical Thomas-Fermi limit. For spin-polarized fermions, there is no short-range 
 contribution to the interaction energy and the virial theorem reported in  Ref.~\cite{goral01} involves the 3D interaction energy 
 associated with the long-range part $-2d^2\, P_2(\cos{\theta})/r^3$ of the 3D dipolar interaction. Their result is consistent with 
 Eq.~\eqref{eq:virialpure} which applies in the presence of a tight confinement along the $z$-direction. Note, however, that the
dipolar contact~\eqref{eq:dipoleadiabatic} is effectively evaluated with $g(r)\equiv 1$ in Ref.~\cite{goral01}. Due to trap-average,  
the $1/r^3$-divergence of the dipole potential~\eqref{eq:Vd} at short distances is removed and thus the integral is finite despite 
neglecting short-range pair correlations. 

\subsection{Momentum distribution and static structure factor}\label{sec:2e}
 
The short-distance factorization of the many-body wave function~\eqref{eq:manybody} also determines the high-momentum tails of 
various correlation functions, such as the momentum distribution and the static structure factor. Specifically, the momentum distribution 
is given by the Fourier transform of the one-particle density matrix:
 \begin{align}
n({\bf q}) &= n \int d({\bf r}_a, {\bf r}_b, {\bf r}_2, \ldots, {\bf r}_N) \, e^{- i {\bf q} \cdot ({\bf r}_a-{\bf r}_b)} \nonumber \\*
& \Psi^*({\bf r}_a, {\bf r}_2, \ldots, {\bf r}_N) \Psi({\bf r}_b, {\bf r}_2, \ldots, {\bf r}_N) ,\label{eq:momentum}
\end{align}
which is a dimensionless quantity that is normalized to the density via $\int_q n({\bf q}) =n$. The dominant singularity at short distances 
arises from configurations in which both coordinates ${\bf r}_a$ and ${\bf r}_b$ approach any of the other $N-1$ integration coordinates 
simultaneously. Using Eq.~\eqref{eq:manybody}, this gives rise to the high-momentum behavior
\begin{align}
\lim_{q\to\infty} n({\bf q}) &= \Bigl| \int d{\bf r} \, e^{-i {\bf q} \cdot {\bf r}} \frac{\phi(r)}{2\pi} \Bigr|^2 {\cal C} . \label{eq:highmomentum}
\end{align}
\begin{figure}[t]
\scalebox{0.8}{\includegraphics{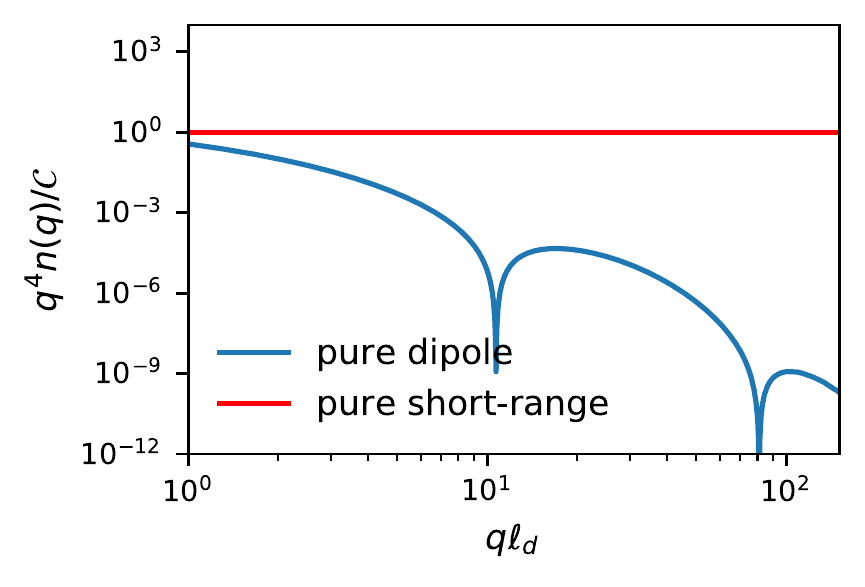}}
\caption{Asymptotic high-momentum tail of the momentum distribution for a pure dipolar interaction as extracted from Eq.~\eqref{eq:highmomentum}.}
\label{fig:2}
\end{figure}
For quantum gases with short-range interactions, the two-body wave function at large momentum is \mbox{$\phi(q)=2\pi/q^2$}. 
The momentum distribution thus exhibits a power-law ${\cal C}/q^4$ tail which in fact holds in any 
space dimension~\cite{tan08b,zhang09,barth11,valiente11,valiente12,werner12a,werner12b}. 
In the presence of dipolar interactions, this behavior remains valid for momenta $q\ell_d \ll 1$, where $\phi(r)$ may be replaced by 
the form $\phi(r) = \ln (r/a_2)$ valid in the regime $\ell_d\ll r\ll 1/\sqrt{n}$. For a pure dipolar gas, $\phi(r)$ becomes exponentially 
small at distances $r<\ell_d$. The momentum distribution is then exponentially suppressed at high momenta as well. Remarkably, 
it exhibits a nontrivial oscillatory structure arising from the Fourier transform of the modified Bessel function $K_0$ in Eq.~\eqref{eq:2bodywf}. 
This is shown in Fig.~\ref{fig:2} where the effective strength of the $1/q^4$ tail at high-momentum for the pure dipolar gas is depicted 
as a function of the dimensionless momentum $q\ell_d$ in a double-logarithmic plot. It exhibits a crossover from a power-law tail in 
the regime $\sqrt{n} \ll q \ll 1/\ell_d$ to exponential suppression for $q\ell_d\gg 1$. In practice, an observation of this peculiar behavior 
requires dipolar lengths $\ell_d$ of the order $\mu$m, which is significantly larger than the values that have been realized so far. 
In the presence of an additional short-range interaction, the exponential divergence of the two-body wave function~\eqref{eq:2bodywf} 
at short distances leads to a high-momentum tail that depends strongly on a cutoff. In contrast to the thermodynamics, the behavior of 
the momentum distribution at large wave vectors is then no longer universal.  
 
An analogous crossover is seen in the Fourier transform of the pair distribution function, which determines the static structure factor~\eqref{eq:defSq}. 
The short-distance result for the pair distribution function in Eq.~\eqref{eq:shortrangegr} implies the large-momentum behavior
\begin{align}
\lim_{q\to\infty} [S({\bf q}) - 1] &= \frac{{\cal C}}{(2 \pi)^2 n} \biggl(\int d{\bf r} \, e^{-i{\bf q} \cdot {\bf r}} |\phi(r)|^2\biggr) . 
\label{eq:statichigh}
\end{align}
In an intermediate momentum region $\sqrt{n} \ll q \ll 1/\ell_d$, the logarithmic dependence of the pair distribution function gives rise to a power-law momentum tail
\begin{align}
S({\bf q}) - 1 = \frac{{\cal C}_{}}{4 \pi n q^2} \ln \frac{q a_{2} e^{\gamma_E}}{2} + \ldots , \label{eq:Sqtail}
\end{align}
which is modified by a logarithmic factor. For even larger momenta $q \gg 1/\ell_d$, this tail will be exponentially suppressed. 
The domain of validity $\sqrt{n} \ll q \ll 1/\ell_d$ 
for the expression~\eqref{eq:Sqtail} shows that the power-law decay is accessible only in the low-density regime $\sqrt{n}\ell_d \ll 1$, 
where the details of the long-range dipole potential are not important.  The power-law tail~\eqref{eq:Sqtail} predicts a 
negative correction below $q < 2 e^{-\gamma_E}/a_{2}$ while above that, the structure factor is larger than unity. Thus, a maximum appears
at $\bar{q} = 2 e^{1/2-\gamma_E}/a_{2}$ which equals $\bar{q} \approx 0.58/\ell_d$ in the case of a pure dipolar interaction. Such a non-monotonic 
behavior has been observed in numerical calculations of the static structure factor of pure dipolar systems by~\textcite{astrakharchik07} close to the transition to a crystalline 
phase at high density $\sqrt{n}\ell_d \gtrsim 20$. This will be discussed in more detail in the following. 

 \subsection{Numerical values of the dipolar contact for pure dipolar interactions}\label{sec:2f}

\begin{figure}[t]
\scalebox{0.8}{\includegraphics{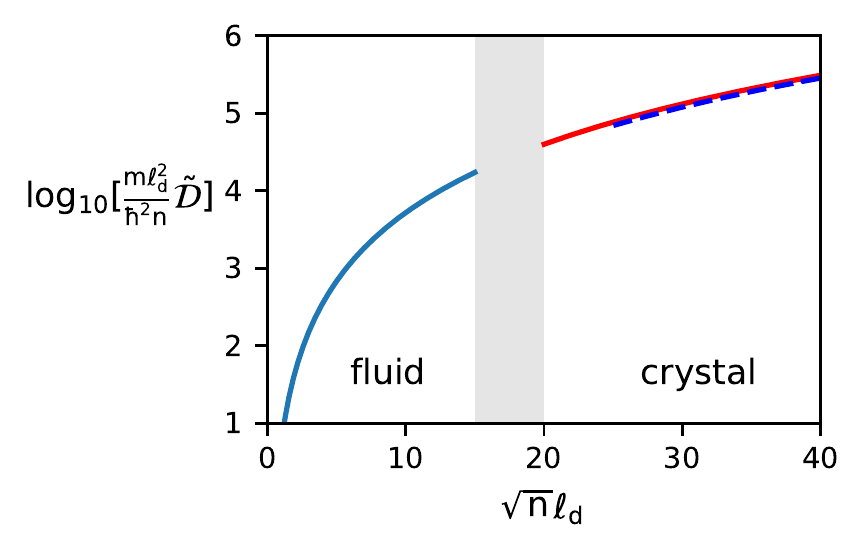}}
\caption{Contact of a pure dipolar gas as extracted from the QMC calculations in Ref.~\cite{astrakharchik07} as a function of the dimensionless ratio of 
dipolar length and interparticle separation $\sqrt{n}\ell_d$. The blue line indicates the contact in the fluid phase and the red line in the crystalline phase, 
with an intermediate transition region between the two phases shown in gray. The exact result for the high-density limit is indicated by the blue dashed line.}
\label{fig:3}
\end{figure}

The result in Eq.~\eqref{eq:dcontactpert} for the dipolar contact $\cal D$ only covers the limit of ultra-low densities. In order to determine the 
contact quantitatively over a wider range --- at least for a system with pure dipolar interactions, where $\cal D\to \tilde{\cal D}$ --- 
we apply the adiabatic derivative~\eqref{eq:dipoleadiabatic} to the numerical results obtained in Ref.~\cite{astrakharchik07}.  As shown 
in Fig.~\ref{fig:3}, the dimensionless dipolar contact increases monotonically with the ratio of the dipolar length $\ell_d$ and the average 
interparticle spacing, ranging over almost six orders of magnitude in a relevant range of densities.

 In the low-density limit $\sqrt{n}\ell_d \ll 1$,  the contact~\eqref{eq:puredipole} follows the logarithmic dependence~\eqref{eq:contactpert} 
derived in Sec.~\ref{sec:2c}. In the opposite high-density limit $\sqrt{n}\ell_d\gtrsim 20$, the system forms a regular triangular lattice. 
The asymptotic dependence of the dipolar contact is then determined by the purely classical energy density of the crystal:
\begin{align}
\tilde{\cal D}_{\rm crystal} &= \frac{n}{2} \sum_{{\bf R} \neq 0} \frac{d^2}{|{\bf R}|^3} = \frac{\hbar^2 n}{m \ell_d^2} (n\ell_d^2)^{3/2} 
\times \frac{1}{2} \sum_{{\bf R} \neq 0} \frac{1}{n^{3/2} |{\bf R}|^3} . \label{eq:highdensity}
\end{align}
The sum runs over the sites ${\bf R}$ of a triangular lattice and the factor $1/2$ avoids double-counting. For a triangular lattice, 
which is the configuration with the lowest ground state energy for purely repulsive interactions, the last numerical factor is $4.4462$, 
which agrees with the corresponding constant $a_1=4.43(1)$ obtained numerically in Ref.~\cite{astrakharchik07}. Note that for a given 
density and strength $d^2$ of the dipolar interaction, $\tilde{\cal D}_{\rm crystal}$ is independent of $\hbar$. Quantum corrections to 
this purely classical energy arise from the zero point motion of the crystal. This leads to an additional contribution to the energy per 
particle of order $\tilde{\varepsilon}_{\rm phon}\simeq \hbar c_s/b \sim n^{5/4}$ because the sound velocity 
scales like $c_s\sim n^{3/4}$, while the lattice constant decreases like $b\sim 1/\sqrt{n}$ with density. The resulting quantum 
corrections to $\tilde{D}_{\rm crystal}$, which are of order ${\cal O}(n (n\ell_d)^{5/4})$, are quite small, however.  Indeed, as shown 
in Fig.~\ref{fig:3}, the full result hardly differs from the dashed blue line representing the contribution~\eqref{eq:highdensity} 
even in the regime close to the transition. 

In the context of the fluid-to-crystal transition found numerically in Refs.~\cite{buechler07,astrakharchik07}, two points merit further discussion. 
First of all,  in two dimensions and in the presence of dipolar or even longer-range interactions, a direct first order transition 
from a fluid to a crystalline phase has been excluded by Spivak and Kivelson on quite general grounds~\cite{spivak04,spivak06}. 
Indeed, such a transition requires a positive value of the surface tension. For interactions decaying like $V(r)\sim 1/r^n$ at large
distances with $n\leq 3$, however, the fluctuation contributions to the surface tension in $d=2$
become negative for large domain sizes~\cite{spivak06}. As a result, one expects an inhomogeneous stripe or micro-emulsion phase 
intervening between the fluid and crystalline ground states. Such phases have been predicted for fast rotating gases in the presence 
of dipolar interactions by Cooper {\it et al.}~\cite{cooper05, komineas07}. In a non-rotating situation, where the
projection to the lowest Landau level does not apply, they are difficult to resolve in numerical simulations, however, because 
in the special case of dipolar interactions the characteristic domain sizes are expected to be larger than the microscopic length 
scales $\ell_d$ or $1/\sqrt{n}$ by an exponentially large factor. 
To account for the presence of such intermediate phases, in Fig.~\ref{fig:3} we have left open a finite interval in the vicinity of
the critical dimensionless coupling $\sqrt{n}\ell_d\simeq 18$. 

As a second point, we note that starting from a homogeneous fluid, the point of instability towards phases with a nontrivial modulation 
of the density may be inferred from an empirical criterion that only involves knowledge of the static structure factor.  In classical liquids, 
this is known as the Hansen-Verlet criterion. It states that freezing appears when the dominant first peak of the static structure factor 
reaches a critical value $S(q_0)=2.85$~\cite{hansen69}. As discussed by~\textcite{babadi13}, a modified version of this criterion turns 
out to determine the limit of stability also for various two-dimensional quantum fluids at zero temperature. Since configurations with a
strongly inhomogeneous density are suppressed in quantum mechanics, the associated critical value $S(q_0)$ is substantially lower than 
the classical Hansen-Verlet value. Surprisingly, it does not change much with particle statistics or the specific form of the repulsive interactions. 
In the particular case of Bose fluids with dipolar interactions, the 
value extracted from the numerical results in Ref.~\cite{astrakharchik07} is $S(q_0)\simeq 1.7$~\cite{babadi13} (an even smaller value 
$S(q_0)\simeq 1.4$ applies for dominantly repulsive Bose fluids in $d=3$ like $^4$He~\cite{kalos74}). In the following section, we will show 
that, at least for quasi-two-dimensional dipolar gases, the Hansen-Verlet criterion essentially coincides with the appearance of a roton 
minimum in the excitation spectrum. With increasing strength of the dipolar interaction, the superfluid therefore becomes unstable towards 
a ground state with a density wave rather than staying in a homogeneous phase with a well-defined roton minimum, a scenario that also 
appears to be realized for dipolar gases in a cigar-shaped trap~\cite{petter19,hertkorn20}.  

\section{Exact relations for quasi two-dimensional dipolar gases}\label{sec:3}

The results derived in the previous section apply for dipolar gases in the limit $R_e, l_z \ll a_2, \ell_d$, where the solution of the two-body 
problem takes the form~\eqref{eq:2bodywf} appropriate for scattering in two dimensions. In practice, such a strong confinement has not 
yet been reached. Indeed, typical dipolar lengths are below $100\, a_B$ and are thus much smaller than the transverse confinement lengths 
on the order of $l_z\simeq 0.5\,\mu$m. In such a case, the short-distance behavior is determined by a solution of the full 
three-dimensional Schr\"odinger equation in the presence of both a short-range and the dipolar potential:
\begin{align}
V({\bf r}) &= V_{\rm sr}({\bf r}) + \frac{d^2}{r^3} \biggl(1 - \frac{3z^2}{r^2}\biggr)  - \frac{8\pi d^2}{3} \delta^{(3)}({\bf r})\, .\label{eq:3Ddipole}
\end{align}
Here, the attractive delta-function term is a dipolar contribution to the short-range interaction which --- in contrast to the contribution $V_{\rm sr}({\bf r})$ --- 
is known explicitely in analytical terms. It is specific to the case of magnetic point dipoles and arises from the short distance behavior 
of the magnetic field whose space integral over a sphere of arbitrary small radius must yield the enclosed dipole moment, ensuring 
that ${{\rm div} \, \mathbf{B}\equiv 0}$ holds globally~\cite{griffiths82}.  

As discussed in the introduction, the associated scattering problem even at low energies cannot be reduced to $s$-wave 
interactions only and thus is not universal. Nevertheless, a number of exact results can be obtained for the weakly-confined 
dipolar gas, where the motion along the confined $z$-direction is restricted to the lowest transverse single-particle eigenstate. 
This requires the chemical potential to obey the condition $\mu\ll\hbar\omega_z$, which excludes a high-density crystalline 
phase as discussed in the previous section. We note that the assumption $\mu\ll\hbar\omega_z$ even for the tightly confined 
directions is not obeyed for dipolar gases in cigar-shaped traps realized so far~\cite{boettcher19,tanzi19,chomaz19}. Nevertheless,
a number of features like the absence of a homogeneous superfluid with a well-defined roton minimum and the finite maximum value 
of the static structure factor right at the transition point are common with those of our two-dimensional model system studied in the following. 
On a qualitative level, this can be understood by the fact that the presence of excited states for the transverse motion, which may be
due to either thermal effects or interactions, may be accounted for by a renormalization of the dimensionless coupling constant $\tilde{g}_2$  
introduced below, see, for example, Ref.~\cite{hadzibabic08}.

\subsection{Effective dipole interaction and stability}\label{sec:3o}

The contribution to the effective two-body interaction $V_{\rm dd}({\bf q})$ that results from projecting the three-dimensional 
dipolar interaction~\eqref{eq:3Ddipole} onto the lowest transverse oscillator level is~\cite{fischer06, komineas07}
\begin{align}
V_{\rm dd}({\bf q}) &= - g_2^{\rm dd} \sqrt{\frac{\pi}{2}} (ql_z) e^{q^2l_z^2/2} {\rm erfc}(\frac{ql_z}{\sqrt{2}})\, .  \label{eq:projecteddipole}
\end{align}
Here, following the notation in Ref.~\cite{bloch08}, we have introduced a coupling constant $g_2^{\rm dd}=(\hbar^2/m) \tilde{g}_2^{\rm dd}$ 
with a dimensionless factor $\tilde{g}_2^{\rm dd}=\sqrt{8\pi} \ell_d/l_z$ for dipolar interactions, which is much less than one in practice. 
The effective interaction $V_{\rm dd}({\bf q}) $ is always negative and approaches the constant value $-g_2^{\rm dd}$ in the limit $ql_z \gg 1$. 
In physical terms, this describes attractive head-to-tail collisions between aligned dipoles with an effective 3D scattering length $-\ell_d$.  
In the opposite limit  $ql_z \ll 1$, the projected dipolar interaction $V_{\rm dd}({\bf q})\sim - 2\pi d^2 q$ 
vanishes with a linear slope, reflecting the repulsive $d^2/r^3$ potential at distances much larger than $l_z$. 
 
The total momentum-dependent interaction $V_{\rm tot}({\bf q})=g_2+V_{\rm dd}({\bf q})$, which arises from the combination of a 
short-range part described by an associated scattering length $a_s$ via $g_2=(\hbar^2/m) \sqrt{8\pi} a_s/l_z$~\cite{bloch08} 
and the magnetic dipolar potential gives rise to a thermodynamically stable low-density gas provided that $a_s>0$. Here, stability 
is understood in the minimal sense that the density response function $\chi({\bf q})$ which describes the change in energy
\begin{equation}
E[\{\delta n_q\}]=E_0 + \frac{1}{2}\int_q \chi^{-1}({\bf q})\,|\delta n_q |^2+\ldots
\label{eq:chi(q)}
\end{equation}
associated with small fluctuations $\delta n_q$ around a homogeneous fluid state is positive in the limit $q\to 0$, where
 $\chi({\bf q})\to\tilde{\kappa}=\partial n/\partial \mu$. In physical terms, $\tilde{\kappa}>0$ is guaranteed by a positive value of the 
 effective scattering length for head-to-head collisions between aligned dipoles. It is important to note that this is a weaker condition 
 compared to the case without a confining potential, where the effective scattering length 
$a_s^{\rm eff}=a_s-\ell_d$ for head-to-tail collisions must be positive~\cite{lima11}. For quasi-2D systems, $a_s^{\rm eff}$ may become 
negative despite overall stability. It is the presence of attractively interacting dipoles in a weakly confined configuration with $l_z\gg\ell_d$ 
that opens the possibility for an instability of the homogeneous superfluid into phases with spatial order. The resulting supersolid phase
arises through the formation of a mass density wave with many particles per unit cell. Such phases have been termed cluster crystals 
and they arise generically for interactions whose Fourier transform is negative in a range near the ordering wave-vector $q_0$~\cite{rossi11}.

\begin{figure}[t]
\scalebox{0.8}{\includegraphics{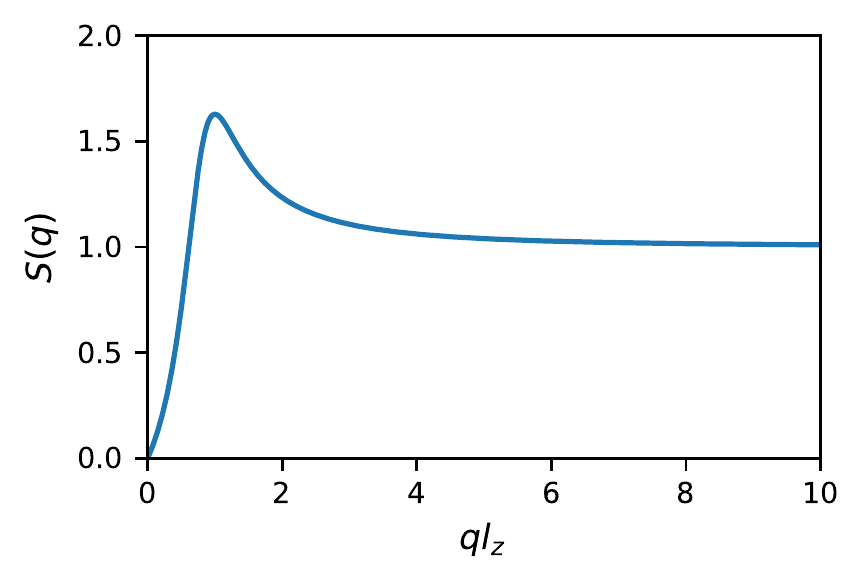}}
\caption{Static structure factor of a weakly interacting Bose gas as predicted from Bogoliubov theory. 
Parameters are chosen such that a roton minimum just appears in the excitation spectrum and the 
dominant peak in the static structure factor at $ql_z\simeq 1.3$ has reached the critical Hansen-Verlet value $1.7$.}
\label{fig:4}
\end{figure}

 \subsection{Static structure factor and Hansen-Verlet criterion}\label{sec:3b}

 In the following, we consider the Hansen-Verlet criterion for weakly confined dipolar gases in two dimensions and show that the appearance of a roton minimum essentially coincides with the point where the homogeneous fluid becomes unstable towards a state with a periodic density-wave. This result indicates that the roton instability, 
 where the excitation gap vanishes at some finite momentum $q_0$, is preempted by a transition to a phase with a finite density modulation. 
It is important to note that, in contrast to the transition to an incompressible, commensurate crystal 
in the tightly-confined limit discussed in the preceding section, the validity of a Hansen-Verlet criterion 
for transitions into a phase with broken translation invariance in the presence of long-range and partially attractive 
interactions has not been studied before and is thus not a priori evident. In particular, the criterion does not fix the nature of the spatial order 
beyond the instability nor does it provide a microscopic description of the underlying first-order quantum phase transition.
 
\begin{figure}[t]
\scalebox{0.8}{\includegraphics{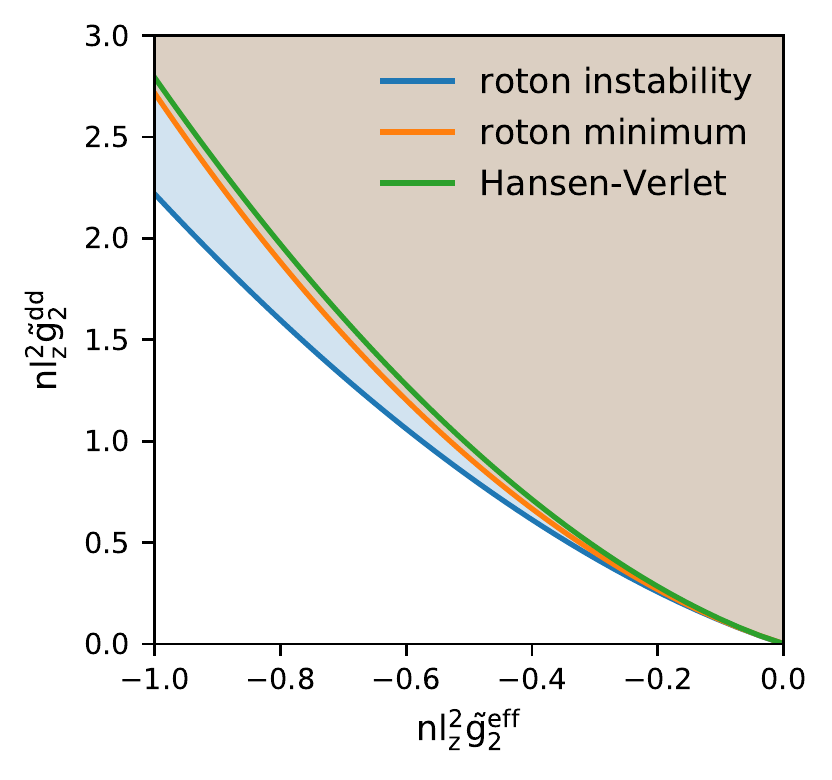}}
\caption{Stability diagram of the weakly confined dipolar Bose gas as obtained within a Bogoliubov approximation. The blue line marks 
the region in parameter space in which the gas is stable (indicated by the blue and brown shaded regions). The orange line shows the 
onset of the roton minimum, and the green line marks the Hansen-Verlet criterion.}
\label{fig:5}
\end{figure}

As noted above, for purely repulsive interactions, the Hansen-Verlet criterion states that the dominant peak in 
the static structure factor reaches a critical value $S(q_0)=\mathcal{O}(1)$ of order one at the transition to a phase 
with an inhomogeneous density, where $\mathcal{O}(1)\simeq 1.7$ in the specific case of $1/r^3$ interactions in two dimensions.
The relevance of this criterion for weakly confined dipolar gases in the limit  $\mu\ll\hbar\omega_z$ can be tested easily at the 
level of a Bogoliubov approximation by noting that the resulting static structure factor
\begin{align}
S_{\rm Bog}({\bf q}) &= \left[ 1 + 2n_0\,V_{\rm tot}({\bf q})/\varepsilon_q\right]^{-1/2} \label{eq:Sqbog}
\end{align}
is completely determined by the effective interaction and the condensate density $n_0$ ($\varepsilon_q=\hbar^2 q^2/2m$ is the single particle energy). 
Based on the expression~\eqref{eq:projecteddipole} for the momentum dependent interaction, Fig.~\ref{fig:4} shows the static structure factor 
at a dimensionless dipolar interaction strength $n_0 l_z^2 \tilde{g}_2^{\rm dd} = 1$  and a negative effective short-range interaction  
$n_0 l_z^2 \tilde{g}_2^{\rm eff} = -0.5$. This parameter regime corresponds to the onset of the roton minimum in the excitation spectrum 
which --- within Bogoliubov theory --- is given by the single-mode expression $E_q=\varepsilon_q/S(q)$~\cite{blakie12}. In Fig.~\ref{fig:5}, we show the associated 
stability diagram. Here, the blue line marks the roton instability, where the excitation energy reaches zero at finite momentum $q_0$ due to a 
formally divergent value of the static structure factor. Remarkably, 
the line where the excitation spectrum $E_q$ starts to develop a roton minimum (orange line) essentially coincides with the Hansen-Verlet 
criterion $S(q_0)\simeq 1.7$ (green line). This suggests that --- in contrast to the case of $^4$He --- 
a homogeneous superfluid with a well-developed roton minimum exists at most within a small range of parameter values: near the point where the roton 
minimum starts to develop, a first-order transition to a supersolid appears which exhibits a mass-density wave but still retains long range phase coherence. 

Remarkably, this scenario for the transition to a supersolid state is observed experimentally with dipolar gases in a cigar-shaped trap, where the chemical potential is larger than the 
zero-point energy even along the two tightly-confined directions. The expression~\eqref{eq:projecteddipole} for the momentum-dependent 
interaction therefore does not apply quantitatively. Nevertheless,  the Bragg scattering data by~\textcite{petter19} show that upon lowering 
the short-range scattering length towards a critical value of order ${\cal O}(\ell_d)$, only a rather shallow minimum develops in the excitation 
spectrum near $ql_z\simeq 1.3$ before the system undergoes a transition to a supersolid phase with a finite density modulation along the 
axial direction.

A detailed analysis of the superfluid-to-supersolid transition has been achieved in the recent measurements 
by~\textcite{hertkorn20} of the structure factor of a dipolar gas of dysprosium in a cigar-shaped trap.  
By averaging around $200$ in situ images of the atomic density, the finite-temperature static structure factor 
\mbox{$S(\mathbf{q},T)=\langle |\delta n_q |^2\rangle(T)/N$} is inferred from the observed shot-to-shot density fluctuations $\delta n_q$. 
Due to the cigar-shaped trap, the fluctuations are strongest along the axial direction, with the dominant peak shifting towards larger values of 
the longitudinal momentum as the transition to a supersolid phase at a critical value of the short-range scattering length is approached. 
At the relevant temperature $T\simeq 20\, {\rm nK}$ of the experiment, the maximum peak value of around $S(q_0,T)\simeq 260$ 
appears at a wave vector $q_0\simeq 2\pi\cdot 0.29\, \mu{\rm m}^{-1}$~\cite{hertkorn20}. To compare with the Hansen-Verlet criterion, an 
estimate of the critical height of the dominant peak at zero temperature may be obtained by using the Bogoliubov approximation 
\begin{align}
S_{\rm Bog}({\bf q},T) &= S_{\rm Bog}({\bf q})\,\coth{\frac{\beta E_q}{2}}  \label{eq:SqT}
\end{align}
with the additional assumption that the dispersion $E_q$ does not depend on temperature. The thermal factor near 
the maximum of the structure factor is in the range $50-100$, which corresponds to a critical peak height at zero temperature 
in the range between $2.6$ and $5.2$. This is larger than the value $1.7$ obtained by applying the Hansen-Verlet criterion in
the case of the tightly confined and  purely repulsive dipolar gas in two dimensions. Given that a number of assumptions enter into 
the extrapolation of the experimental data to zero temperature, this is not an unreasonable deviation. In addition, it should be 
emphasized that the maximum peak height $S_{\rm max}(q_0)$ of the zero temperature structure factor in the quantum version of 
the Hansen-Verlet criterion is not universal.  Instead, in general the number depends on dimensionality and 
the detailed form of the interaction. What the Hansen-Verlet criterion says, however, is that $S_{\rm max}(q_0)$ is a constant of order one
for any first order quantum phase transition between a fluid phase and one with a periodic modulation in the density.
This appears to be the case also for the superfluid-to-supersolid phase transition of dipolar gases in a cigar-shaped 
trap. Indeed, experimentally, after crossing the transition at the critical value of the short-range scattering length,
the dominant peak in the static structure factor is observed to decrease~\cite{hertkorn20} within the supersolid phase,
where the spacing of the droplets is close to $2\pi/q_0$. The roton-instability predicted by the Bogoliubov approximation, 
where the static structure factor diverges at some finite momentum $q_0$, is thus preempted by the spatially ordered supersolid phase.  

\subsection{Static structure factor beyond Bogoliubov theory}\label{sec:3c}

In order to determine to which extent features in the static structure factor provide information about the nature of the instability towards 
inhomogeneous phases that remain valid beyond the Bogoliubov approximation, we first note that the associated high-momentum tail
\begin{align}
\lim_{ql_z \gg 1} S_{\rm Bog}(q) &= 1 - \frac{4\sqrt{2 \pi} n_{2} a_s^{\rm eff}}{l_z q^2} + \ldots \label{eq:bogoliubov}
\end{align}
is determined by the effective scattering length $a_s^{\rm eff}=a_s-\ell_d$. The trivial asymptotic limit $S(q)=1$ is thus approached from 
above provided $a_s^{\rm eff}<0$ is negative. It turns out, however, that the Bogoliubov approximation does not account for the correct
asymptotic behavior of the static structure factor at large momentum. In fact, the tail of $S(q)$ for large in-plane momenta ${\bf q}$ probes 
dipoles at lateral separations $\boldsymbol{\rho}$ that are much smaller than the vertical displacement in the transverse direction. 
This can be seen from the definition of the static structure factor in terms of the pair distribution function
\begin{align}
S({\bf q}) &= 1 + n_3 \int d\boldsymbol{\rho} \, e^{-i{\bf q} \cdot \boldsymbol{\rho}} \int dz (g(\boldsymbol{\rho}, z) - 1) . \label{eq:defSq3D}
\end{align}
In the limit $ql_z\gg 1$, the dominant contribution comes from dipoles with lateral separation $|\boldsymbol{\rho}|\ll l_z$, which is 
averaged over the direction $z$ of the dipoles. Quite different from the fluid-to-crystal transition discussed in section II.F, which is driven by purely repulsive 
interactions, the dipoles now may interact attractively. In particular, for distances below $l_z$ the transverse confinement is not felt and the 
scattering problem is of a three-dimensional nature with an effective negative scattering length $a_s^{\rm eff}<0$, which describes the strength 
of head-to-tail collisions. For length scales considerably larger than the dipole length $\ell_d$, the interaction is well described by 
a pseudopotential description based on the standard Bethe-Peierls boundary condition. 
As a result, the short-distance behavior of the pair distribution function is of the form discussed by~\textcite{tan08a} or~\textcite{zhang09}
\begin{align}
g(\boldsymbol{\rho}, z) &\sim \biggl(\frac{1}{r} - \frac{1}{a_s^{\rm eff}}\biggr)^2 .
\end{align}
Performing the Fourier transform in Eq.~\eqref{eq:defSq3D}, the structure factor 
\begin{align}
S(q) - 1&\sim \frac{1}{8 n_3 q} \biggl(1 - \frac{4}{\pi q a_s^{\rm eff}} + \ldots \biggr) . \label{eq:SqTan}
\end{align}
exhibits a high-momentum tail analogous to the one obtained for Bose gases in three-dimensions~\cite{hofmann17}.
This result holds for wave vectors larger than the inverse oscillator length $1/l_z$,  yet smaller than inverse dipole length $1/\ell_d$, 
beyond which the details of the interaction at short distances become important. With $l_z/\ell_d \simeq 150$ in current experiments~\cite{petter19}, 
this is a broad window. Independent of the sign of $a_s^{\rm eff}$, the static structure factor thus always approaches unity from above as $1/q$. 
For negative values $a_s^{\rm eff} <0$, which is the case relevant to current experiments~\cite{petter19, petter20}, 
also the subleading contribution is positive. As a result, the static structure factor exhibits a monotonic decay from its dominant peak 
at $q_0l_z={\cal O}(1)$ towards the limiting value one, as shown in Fig.~\ref{fig:4}. This is quite different from the situation found with purely 
repulsive interactions, where $S(q)$ exhibits both a minimum and a maximum at wave vectors beyond $\sqrt{n}$, see, for example, Ref.~\cite{astrakharchik07}
and the discussion at the end of the previous section. 

The Bogoliubov approximation~\eqref{eq:bogoliubov}, by contrast, fails to correctly describe the asymptotic form of the static structure 
factor~\eqref{eq:SqTan} and only captures the subleading contribution $\sim 1/q^2$, missing the exact behavior~\eqref{eq:SqTan} that 
always approaches unity from above. A similar situation is also found for Bose gases with pure short-range interactions 
and in the absence of a confinement~\cite{hofmann17}.

\section{Summary}\label{sec:4}

In summary, we have shown that tightly confined dipolar gases admit a universal description that extends those developed by 
Tan and by Zhang and Leggett in the case of short-range interactions. The description is based on only two experimentally tunable 
parameters, the two-dimensional scattering length and the dipolar length scale. The associated adiabatic derivatives of the grand
 canonical potential define a generalized contact parameter and an additional dipolar analog of the contact. These two contact parameters 
 determine thermodynamic relations such as the pressure of a homogenous system as well as the virial theorem in a trapped gas. 
 Explicit results for both contacts have been given for zero temperature in the limit of low densities. 
 In addition, we have discussed the behavior of the momentum distribution $n(q)$ and the static structure factor at large wave vectors. 
 The standard $\mathcal{C}/q^4$-tail in $n(q)$ for short-range interactions is replaced by a more complicated structure, exhibiting
 a characteristic minimum around $q\ell_d\simeq 10$.  
 
The results presented in the first part of this paper apply in the limit of strong transverse confinement, a limit that has not yet been realized
experimentally. In the second part, a number of results of a rather general nature have been derived for dipolar gases in a quasi-two-dimensional 
configuration. In particular, the high-momentum behavior of the static structure factor allows to distinguish the density wave instability in weakly confined 
dipolar gases from those in dense quantum liquids. Specifically, in the former case one expects a monotonic decay 
from the dominant peak in the static structure factor towards the asymptotic value of unity. 
Moreover, it has been shown that the appearance of a roton minimum in the excitation spectrum essentially coincides 
with the point where dipolar gases become unstable towards a density wave instability according to the empirical Hansen-Verlet criterion
for freezing, originally developed for fluid-to-solid transitions with dominantly repulsive short-range interactions. 
With increasing strength of the dipolar interaction, the roton instability predicted within Bogoliubov theory is thus
preempted by a first-order transition to a state with a non-vanishing density modulation. The observation that the Hansen-Verlet criterion apparently describes a number of generic features which are observed in the superfluid-to-supersolid transition of dipolar gases in a cigar-shaped trap is quite remarkable and deserves further investigation.

\begin{acknowledgements}
It is a pleasure to acknowledge a number of helpful comments by G. Astrakharchik, H. L\"owen, D. Petrov,  A. Recati, B. Spivak and M. Zwierlein.
This work is supported by Peterhouse, Cambridge, and Vetenskapsr\aa det (grant number 2020-04239) (J.H.).
\end{acknowledgements}

\appendix
\section{Two-body scattering in soft core plus power-law potentials}\label{app:B}

It is instructive to discuss an example for the two-body scattering problem with a power-law potential together with a non-universal short-range part. 
In this appendix, we consider the analytically soluble case of scattering from a combined potential-well plus power-law interaction
\begin{align}
V(r) &=
\begin{cases}
- V_0 & r<R \\[1ex]
\dfrac{d^2}{r^3} & r\geq R
\end{cases} \label{eq:potential}
\end{align}
with $V_0>0$, and link the microscopic parameters $R$ and $V_0$ to the universal scattering parameters $a_2$ and $\ell_d=(md^2/\hbar^2)^{1/2}$. The total scattering length $a_2$ has a shape resonance whenever parameters are chosen such that the low-energy scattering state can interact with a bound state at threshold.

\begin{figure*}[t]
\subfigure[]{\scalebox{0.7}{\includegraphics{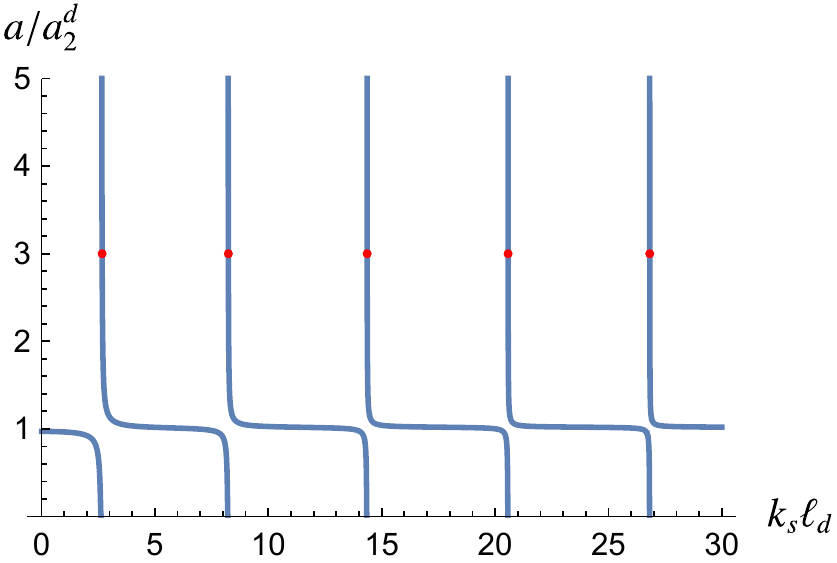}}} \qquad\qquad
\subfigure[]{\scalebox{0.8}{\includegraphics{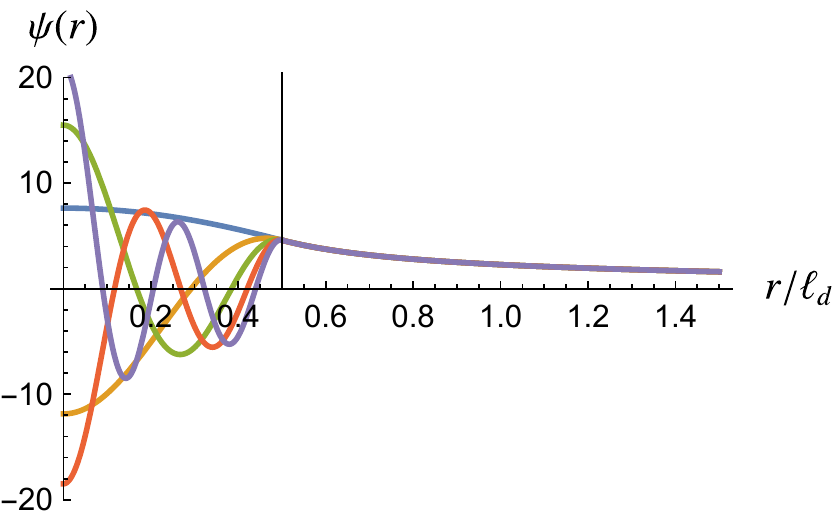}}} 
\caption{(a) Scattering length of a combined dipole and short-range potential with range $R=0.5\ell_d$ as a function of the potential depth $V_0 = \hbar^2 k_s^2/m$. The shape resonance occurs whenever a new bound state appears near threshold. The red dot marks the values at which the scattering length is equal to $a=3a_{dd}$. (b) First five wave functions with scattering length $a=3a_{dd}$. The vertical black lines indicates the range $R=0.5\ell_d$.}
\label{fig:6}
\end{figure*}

The regular solution of the $s$-wave scattering equation~\eqref{eq:dimensionless} in the region $r<R$ reads
\begin{align}
\psi_0^{(1)}(r) &= \sqrt{k'r} \, J_{0}(k'r) \label{eq:psi1f}
\end{align}
with $k' = \sqrt{k^2 + k_s^2}$, where $k_s = \sqrt{2\mu V_0/\hbar^2}$. Outside the potential well, two regions exist in which analytical results for the low-energy scattering can be obtained. First, for $R<r\ll1/k$, we neglect the kinetic term in Eq.~\eqref{eq:dimensionless}, which gives ($\tilde{r} = r/\ell_d$)
\begin{align}
\psi_0^{(2)}(\tilde{r}) &= c_1 \sqrt{\tilde{r}} \, K_{0}\biggl(\sqrt{\frac{4}{\tilde{r}}}\biggr) + c_2 \sqrt{\tilde{r}} \, I_{0}\biggl(\sqrt{\frac{4}{\tilde{r}}}\biggr) , \label{eq:psi2c}
\end{align}
with two matching coefficients $c_1$ and $c_2$. At very large distances $r\gg \ell_d$, by contrast, the dipole interaction is negligible, and the solution of the Schr\"odinger equation is
\begin{align}
\psi_0^{(3)}(r) &= c_1' \sqrt{kr} \, J_{0}(kr) + c_2' \sqrt{kr} \, Y_{0}(kr) . \label{eq:psi3c}
\end{align}
The two limiting solutions~\eqref{eq:psi2c} and~\eqref{eq:psi3c} overlap in a region $R, \ell_d \ll r \ll 1/k$~\cite{ticknor09}. Matching the solutions in this region gives
\begin{align}
c_1' &= \frac{1}{2 \sqrt{k \ell_d}} \Bigl(\ln \frac{2 e^{-2 \gamma_E}}{k \ell_d} - \gamma_E\Bigr) \Bigl(c_1 - \frac{2}{\pi} c_2\Bigr) \\
c_2' &= \frac{\pi}{4 \sqrt{k \ell_d}} c_1 ,
\end{align}
which gives the scattering length in terms of the coefficients $c_1$ and $c_2$:
\begin{align}
a &= e^{2\gamma_E} \ell_d e^{- 2 c_2/c_1} . \label{eq:a1}
\end{align}
Indeed, this identity links the wave function~\eqref{eq:psi2c} with appropriate normalization to the universal two-body wave function~\eqref{eq:2bodywf}. Without the irregular solution $c_2=0$, the scattering length is the dipolar scattering length $a_2^d$ discussed in the introduction. In our model, the coefficients $c_1$ and $c_2$ are, in turn, determined by matching the wave functions at the short-range boundary $r=R$ (defining $\tilde{R} = R/\ell_d$ and $\tilde{k}_s = k_s\ell_d$):
\begin{align}
&c_1 = \nonumber \\
&\frac{\sqrt{\tilde{k}_s} I_1(\sqrt{\tfrac{4}{\tilde{R}}}) J_0(\tilde{k}_s \tilde{R}) - \sqrt{\tilde{k}_s^3 \tilde{R}^3} I_0(\sqrt{\tfrac{4}{\tilde{R}}}) J_1(\tilde{k}_s \tilde{R})}{I_1(\sqrt{\tfrac{4}{\tilde{R}}}) K_0(\sqrt{\tfrac{4}{\tilde{R}}}) + 
   I_0(\sqrt{\tfrac{4}{\tilde{R}}}) K_1(\sqrt{\tfrac{4}{\tilde{R}}})} \\
&c_2 = \nonumber \\
&\frac{\sqrt{\tilde{k}_s } K_1(\sqrt{\tfrac{4}{\tilde{R}}}) J_0(\tilde{k}_s \tilde{R}) + \sqrt{\tilde{k}_s^3 \tilde{R}^3} K_0(\sqrt{\tfrac{4 }{\tilde{R}}}) J_1(\tilde{k}_s \tilde{R})}{I_1(\sqrt{\tfrac{4}{\tilde{R}}}) K_0(\sqrt{\tfrac{4}{\tilde{R}}}) + I_0(\sqrt{\tfrac{4}{\tilde{R}}}) K_1(\sqrt{\tfrac{4}{\tilde{R}}})} .
\end{align}
The full expression for the scattering length is thus:
\begin{widetext}
\begin{align}
a &= e^{2\gamma_E} \ell_d \exp\biggl[- 2 \frac{\sqrt{\tilde{k}_s} K_1(\sqrt{\tfrac{4}{\tilde{R}}}) J_0(\tilde{k}_s \tilde{R}) + \sqrt{\tilde{k}_s^3 \tilde{R}^3} K_0(\sqrt{\tfrac{4}{\tilde{R}}}) J_1(\tilde{k}_s \tilde{R})}{\sqrt{\tilde{k}_s} I_1(\sqrt{\tfrac{4}{\tilde{R}}}) J_0(\tilde{k}_s \tilde{R}) - \sqrt{\tilde{k}_s^3 \tilde{R}^3} I_0(\sqrt{\tfrac{4}{\tilde{R}}}) J_1(\tilde{k}_s \tilde{R})}\biggr] . \label{eq:afull}
\end{align}
\end{widetext}
As a check, consider the limit $\tilde{R} \ll 1$ with $V_0$ fixed, in which the functions $K_0$ and $K_1$ are exponentially suppressed and the functions $I_0$ and $I_1$ are exponentially divergent, with all other terms in Eq.~\eqref{eq:afull} finite. Indeed, this is the limit in which the short-range potential is negligible compared to the dipole interaction, and the expression for the scattering length reduces to the scattering length of the pure dipolar potential, $a_{dd} = e^{2\gamma_E} \ell_d$. In the opposite limit $\tilde{R} \gg 1$ (again keeping $V_0$ fixed), where the dipolar potential is negligible, we reproduce the standard result for the scattering length of a potential well,
\begin{align}
a &=  R \exp\biggl[\frac{J_0(k_s R)}{k_s R J_1(k_s R)}\biggr] . \label{eq:ahc}
\end{align}
This result is derived from Eq.~\eqref{eq:afull} by noting that the function $I_1$ is subleading compared to $K_1$ and $I_0$, which have limits of $\sqrt{\tilde{R}/2}$ and $1$, respectively, while $K_0$ has a logarithmic divergence that changes the prefactor.

Figure~\ref{fig:6}(a) shows the scattering length for one particular potential with $R=0.5\ell_d$ as a function of the scaling variable $k_s\ell_d$ that sets the depth of the potential. Red points in this figure mark the parameters values corresponding to a scattering length $a=3a_{dd}$. Figure~\ref{fig:6}(b) shows the scattering wave functions corresponding to these parameter values. The wave functions take a universal form for $r>R$ but are non-universal below that with a number of nodes that increases with the number of bound states. Note that it is possible to choose the parameters of this potential in such a way that the wave function does not contain any nodes, which might be useful for numerical simulations. From Fig.~\ref{fig:6} it is apparent that with decreasing range $R$, the resonances are more widely spaced and become narrower since the dipolar potential forms an increasingly strong tunneling barrier, until for vanishing $R$ the scattering length is constant and equal to $a_{dd}$, as discussed above. A scaling limit for the potential~\eqref{eq:potential} requires to take $R\to 0$ and $V_0\to\infty$ such that~\eqref{eq:afull} is kept fixed.

\section{Derivation of the adiabatic relations}\label{app:derivation}

In this appendix, we present the derivation of the adiabatic relations~\eqref{eq:adiabatic1} and~\eqref{eq:adiabatic2} using the short-distance 
factorization of the many-body wave function~\eqref{eq:manybody}. We begin by considering two energy eigenstates of the Hamiltonian~\eqref{eq:hamiltonian} 
with different total scattering length and dipole strength, denoted by an index $\alpha$ and $\beta$, respectively: 
$\hat{H}_\alpha | \Psi_\alpha \rangle = E_\alpha | \Psi_\alpha \rangle$, and $\hat{H}_\beta | \Psi_\beta \rangle = E_\beta | \Psi_\beta \rangle$. The difference in energy is
\begin{align}
&(E_\alpha - E_\beta) \langle \Psi_\beta | \Psi_\alpha \rangle = \langle \Psi_\alpha | \hat{H} \Psi_\beta \rangle - \langle \hat{H} \Psi_\alpha | \Psi_\beta \rangle \nonumber \\
&= \int' d({\bf r}_1, {\bf r}_2, {\bf X}) \biggl\{- \frac{\hbar^2}{2m} \sum_{i=1}^N  \Bigl[ \Psi_\alpha^* \nabla^2_{i} \Psi_\beta - \Psi_\beta \nabla^2_{i} \Psi_\alpha^*\Bigr] \nonumber \\
&+ \sum_{i<j}^N \Big[V_\alpha({\bf r}_i - {\bf r}_j) \Psi_\alpha^* \Psi_\beta - V_\beta({\bf r}_i - {\bf r}_j) \Psi_\alpha \Psi_\beta^*\Bigr]\biggr\} . \label{eq:energydifference}
\end{align}
The prime on the integral denotes a restriction to a domain that excludes short-distance regions where two particle coordinates are close to each other, 
$|{\bf r}_i - {\bf r}_j| < \varepsilon$. The hypothesis is that if the system is universal, we are free to exclude this region in Eq.~\eqref{eq:energydifference} 
and then take the limit $\varepsilon \to 0$ such that the result is independent of $\varepsilon$. Applying the divergence theorem at the short-distance boundaries gives
\begin{align}
&(E_\alpha - E_\beta) \langle \Psi_\beta | \Psi_\alpha \rangle = \frac{N(N-1)}{2} \int' d({\bf R}, {\bf X})  \Bigl\{- \frac{2 \pi \hbar^2 \varepsilon}{m}\nonumber \\
&\times \bigl[ \Psi_\alpha^* \tfrac{\partial \Psi_\beta}{\partial r} - \Psi_\beta \tfrac{\partial \Psi_\alpha^*}{\partial r} \bigr]_{{}_{r=\varepsilon}} + \int'_{\bf r} \, 
\big[V_\alpha \Psi_\alpha^* \Psi_\beta - V_\beta \Psi_\alpha \Psi_\beta^*\bigr]\Bigr\} . \label{eq:diff2}
\end{align}
For small variations $\delta a_2 = a_{2,\alpha} - a_{2,\beta}$ and $\delta \ell_d = \ell_{d,\alpha} - \ell_{d,\beta}$, the boundary term in Eq.~\eqref{eq:diff2}
is evaluated using Eq.~\eqref{eq:manybody} along with the relation
\begin{align}
&\varepsilon \bigl[ \phi_\alpha^* \tfrac{\partial \phi_\beta}{\partial r} - \phi_\beta \tfrac{\partial \phi_\alpha^*}{\partial r} \bigr]_{{}_{\varepsilon}} 
= - \frac{\delta a_2}{a_2} + \delta \ell_d 
\int_{\varepsilon}^\infty dr \, \frac{|\phi(r)|^2}{r^2} .
\end{align}
Substituting this result in Eq.~\eqref{eq:diff2} and varying with respect to the universal parameters $\ln a_2$ and $\ln \ell_d$, we obtain the adiabatic 
relations for the energy density $\varepsilon = E/A$ stated in Eqs.~\eqref{eq:adiabatic1} and~\eqref{eq:adiabatic2}.

\section{Universal relations for general power law potentials}\label{app:general}

In this appendix, we show that the results derived in section II may be generalized to repulsive inverse power law interactions 
in two dimensions of the form 
\begin{align}
V_\sigma(r) &= \frac{C_{2+\sigma}}{r^{2+\sigma}} , \label{eq:Vp}
\end{align}
where \mbox{$\sigma>0$} is arbitrary. The length scale of the power-law part is $\ell_{\sigma} = (m C_{2+\sigma}/\hbar^2)^{1/\sigma}$ with an associated scattering length $a_2^{\sigma} = \ell_\sigma \exp[2 (\gamma_E - \ln \sigma)/\sigma]$. The potential~\eqref{eq:Vp} includes several important special cases: The limit $\sigma\to 0$ describes a repulsive scale-invariant 2D generalization of the integrable Calagero-Sutherland-Moser problem in 1D~\cite{landau65,astrakharchik06}, $\sigma=1$ is the dipolar potential discussed in this paper, $\sigma=3$ is a quadrupole potential, $\sigma=4$ is a repulsive van der Waals potential that describes the interaction between Rydberg states~\cite{honer10}, and the limit $\sigma\gg 1$ essentially describes the hard-core limit~\cite{koscik19}. The restriction to $\sigma>0$ ensures proper extensive thermodynamics~\cite{fisher64} and implies that at large distances the potential is subleading compared to the kinetic term, such that the scattering properties are still determined in terms of the characteristic length $\ell_\sigma$ as well as the total scattering length $a_2$ of the combined short-range plus power-law potential. This excludes the case of the Coulomb interaction potential with $\sigma=-1$, where a homogeneous, neutralizing background is needed for stability and a separate set of universal relations has been derived previously by the present authors~\cite{hofmann13}.

The central assumption as before is the separability~\eqref{eq:manybody} of the many-body wave function at short distances with a relative part ($\tilde{r} = r/\ell_\sigma$)
\begin{align}
\phi(r) &= \frac{2}{\sigma} K_{0}\Bigl(\frac{2 \sigma^{-1}}{\tilde{r}^{\sigma/2}}\Bigr) - \ln\Bigl( \frac{a_2}{a_{2}^\sigma}\Bigr) \, I_{0}\Bigl(\frac{2\sigma^{-1}}{ \tilde{r}^{\sigma/2}}\Bigr) ,
\end{align}
which follows from the regular and irregular solution of the two-body scattering problem at low energy. The short-distance behavior of the pair distribution function is still given by the form~\eqref{eq:shortrangegr} with a contact ${\cal C}$ as defined in Eq.~\eqref{eq:contactdensity}. For a pure power-law interaction, the pair distribution function is then exponentially suppressed near the origin as $\exp[-2\sigma^{-1}/\tilde{r}^{\sigma/2}]$, otherwise it diverges as $\exp[2\sigma^{-1}/ \tilde{r}^{\sigma/2}]$. The long-distance asymptotic form~\eqref{eq:grlarger} that depends on the compressibility remains unchanged.

Considering a change in the grand canonical potential,
\begin{align}
d\Omega &= - S dT - P dA - N d\mu + X_a \, d(\ln a_2) + X_\sigma \, d(\ln \ell_\sigma) , \label{eq:defcontacts2}
\end{align}
the corresponding adiabatic relations read
\begin{align}
\frac{X_a}{A} &= \frac{\partial \varepsilon}{\partial (\ln a_2)} \biggr|_{\ell_{\sigma}} = \frac{\hbar^2}{4 \pi m} {\cal C} \\
\frac{X_\sigma}{A} &= \frac{\partial \varepsilon}{\partial (\ln \ell_{\sigma})} \biggr|_{a_2} = \sigma {\cal D}_{{\sigma}} ,
\end{align}
where we define a generalized power-law contact
\begin{align}
{\cal D}_{{\sigma}}  
&= \frac{C_{2+\sigma}}{2} \int d{\bf r} \, \frac{n^2 g(r) - \frac{|\phi(r)|^2}{(2 \pi)^2} {\cal C}}{r^{2+\sigma}} .
\end{align}
Furthermore, we note the pressure relation
\begin{align}
P &= \varepsilon + \frac{\hbar^2 {\cal C}}{8 \pi m} + \frac{\sigma {\cal D}_{\sigma}}{2} 
\end{align}
and the virial theorem
\begin{align}
E &= 2 \langle V_{\rm ext} \rangle - \frac{\hbar^2}{8 \pi m} \int_{\bf R} \, {\cal C}({\bf R}) - \frac{\sigma}{2} \int_{\bf R} \, {\cal D}_{\sigma}
 ({\bf R})  .
\end{align}
As before, for a pure power-law interaction, where the scattering length is proportional to the length scale set by the power-law part, the two contact parameters do not appear independently in the thermodynamic relations. Instead, they involve the interaction energy
\begin{align}
\tilde{\cal D}_{\sigma} &= \frac{\partial \varepsilon}{\partial (\ln \ell_{\sigma})} \biggr|_{a_2=a_{2}^\sigma}
= \frac{C_{2+\sigma}}{2} \int d{\bf r} \, \frac{n^2 g(r) }{r^{2+\sigma}} .
\end{align}
At very low densities $n \ell_{\sigma}^{2} \ll 1$, the explicit form is
\begin{align}
\tilde{\cal D}_{\sigma} &= \frac{\hbar^2}{4 \pi m \sigma} {\cal C}(a_2=a_{2}^{\sigma}) ,
\end{align}
where the low-density contact is given by Eq.~\eqref{eq:contactpert}. At very large densities, we have $\tilde{D}_{\sigma} = \frac{\hbar^2 n}{m \ell_\sigma^2} (n\ell_p^2)^{(2+\sigma)/2} a_\sigma$ with $a_\sigma = 4.4462, 2.8915, 2.3595, 2.07043$ for $\sigma=1,2,3,4$ in a triangular lattice.

\bibliography{bib_dipolar}

\end{document}